\documentstyle[twoside,fleqn,espcrc2]{article}

\title{\bf
Improved Transfer-Matrix Schemes of Phenomenological Renormalization
\thanks{Poster}}

\author{
M.~A.~Yurishchev
\address{\sl
L.~D.~Landau Institute for Theoretical Physics of Russian
Academy of Sciences, 142432 Chernogolovka, Moscow Region,
Russia}
\address{\sl Vasilsursk Laboratory, Radiophysical Research Institute,
606263 Vasilsursk, Nizhny Novgorod Region, Russia}}

\begin{document}

\begin{abstract}
Different phenomenological RG transformations based on scaling
relations for the derivatives of the inverse correlation length
and singular part of the free-energy density are considered.
These transformations are tested on the 2D square
Ising and Potts models as well as on the 3D simple-cubic Ising model.
Variants of RG equations yielding more accurate results than
Nightingale's RG scheme are obtained.
In the 2D case the finite-size equations which give the {\it exact\/}
values of the critical point or the critical exponent are found.
\end{abstract}

\maketitle


\section{Introduction}
\label{sec:Intro}
The phenomenological renormalization-group (RG) method \cite{Nigh7679}
is a powerful tool for the investigation of critical phenomena.
As it is known \cite{dSS81}, phenomenological RG can be
constructed by using not just the correlation length as
it is done in Nigtingale's approach \cite{Nigh7679}, but using any
other quantity with a power-law divergence at criticality.
Binder \cite{B81} suggested a phenomenological renormalization scheme
by using the order-parameter moments (cumulants) which are, on the one
hand, related to the higher susceptibilities and, on the other
hand, immediately suitable for the Monte Carlo simulations.
Recently Itakura \cite{I96} extended Binder's cumulant crossing
method taking linear combination of several different reduced
moments.

In this report, I discuss various RG
transformations which follow from general scaling functional
equations.
These equations are evaluated in terms
of the eigenvalues and eigenvectors of the transfer matrices.
By large transverse sizes of partly finite subsystems, all
those transformations must yield the same results.
However, for the small sizes which normally are used in practice,
different RG equations lead to estimates with distinct
accuracies.
My aim is to find the best strategies of a phenomenological
renormalization group method.
This is especially important for 3D systems.


\section{Phenomenological RG Equations}
\label{sec:PRGE}
Let us write the finite-size scaling equations for the
{\it derivatives\/} of the inverse correlation length $\kappa_L$ and
the singular part of the reduced free-energy density $f^s_L$:
\begin{equation}
   \label{eq:kappa^mn}
   \kappa_L^{(m,n)}(t, h)
   = b^{my_t+ny_h-1}\kappa_{L/b}^{(m,n)}(t', h'),
\end{equation}
\begin{equation}
   \label{eq:f^smn}
   f_L^{s\,(m,n)}(t, h)
   = b^{my_t+ny_h-d}f_{L/b}^{s\,(m,n)}(t', h'),
\end{equation}
where $z^{(m,n)}(x,y)=\partial^{m+n}z/\partial x^m\partial y^n$
($z$ is $\kappa_L$ or $f_L^s$), $t=K-K_c$ is the deviation from
critical coupling, $h$ is a normalized external field, $y_t$ and
$y_h$ are, respectively, thermal and magnetic
critical exponents of the system, $d$ is the space
dimensionality, $L$ is a characteristic size of a subsystem and
$b=L/L'$ is the rescaling factor.


In the phenomenological approach proposed by Nightingale
\cite{Nigh7679}, eq.(\ref{eq:kappa^mn}) with $m=n=0$ is
combined with the ordinary expression for the inverse correlation
length
\begin{equation}
   \label{eq:kappa_L}
   \kappa_L=\ln(\lambda_1^{(L)}/\lambda_2^{(L)}),
\end{equation}
in which $\lambda_1^{(L)}$ and $\lambda_2^{(L)}$ are the largest
and second-largest eigenvalues, respectively, of the associated
transfer matrix.
In the absence of a symmetry breaking field, the critical coupling
$K_c$ is estimated from the equation
\begin{equation}
   \label{eq:kappa}
   L\kappa_L(K_c)=(L-1)\kappa_{L-1}(K_c).
\end{equation}
In writing this equation, one sets $L'=L-1$.

Another possible way to produce a phenomenological renormalization
group is obtained by using eq.(\ref{eq:f^smn}) with $m=n=0$.
The fixed point is given by the relation
\begin{equation}
   \label{eq:fs}
   L^df_L^s(K_c)=(L-1)^df_{L-1}^s(K_c).
\end{equation}
The dimensionless free-energy density, $f_L=f_\infty + f_L^s$,
of a subsystem $L^{d-1}\times\infty$ is calculated by the formula
\begin{equation}
   \label{eq:f_L}
   f_L=L^{1-d}\ln\lambda_1^{(L)}
\end{equation}
and the ``background'' $f_\infty$ is introduced as an extra parameter.

Besides eqs.(\ref{eq:kappa}) and (\ref{eq:fs}), in this paper
I also consider the following RG equations (resulting from the
relations (\ref{eq:kappa^mn}) and (\ref{eq:f^smn})):
\begin{equation}
   \label{eq:chi4}
   \frac{\chi_L^{(4)}}{L^d\chi_L^2}\Big|_{K_c}=
   \frac{\chi_{L-1}^{(4)}}{(L-1)^d\chi_{L-1}^2}\Big|_{K_c},
\end{equation}
where $\chi_L=\partial^2f_L/\partial h^2|_{h=0}=f_L^{s\,(0,2)}(K,0)$
is the zero-field susceptibility and
$\chi_L^{(4)}=\partial^4f_L/\partial h^4|_{h=0}=f_L^{s\,(0,4)}(K,0)$
is a nonlinear susceptibility (eq.(\ref{eq:chi4}) corresponds to
Binder's phenomenological renormalization group);
\begin{equation}
   \label{eq:kappa1}
   L^{2-d}(\kappa_L^{(1)})^2/\chi_L=
   (L-1)^{2-d}(\kappa_{L-1}^{(1)})^2/\chi_{L-1};
\end{equation}
\begin{equation}
   \label{eq:kappa2}
   L^{1-d}\kappa_L^{(2)}/\chi_L=
   (L-1)^{1-d}\kappa_{L-1}^{(2)}/\chi_{L-1};
\end{equation}
\begin{equation}
   \label{eq:kappa4}
   L^{1-2d}\kappa_L^{(4)}/\chi_L^2=
   (L-1)^{1-2d}\kappa_{L-1}^{(4)}/\chi_{L-1}^2.
\end{equation}
Here,
$\kappa_L^{(n)}=\partial^n\kappa_L/\partial
h^n|_{h=0}=\kappa_L^{(0,n)}(K,0)$.
Expressions for the derivatives of the inverse correlation length
and the free energy with respect to $h$ in terms of eigenvalues and
eigenvectors of the transfer matrix are available in \cite{Yu9497}.


\section{Results and discussions}
\label{sec:RD}
To represent numerical data in tables, eqs.(\ref{eq:kappa}),
(\ref{eq:fs}), (\ref{eq:chi4}), (\ref{eq:kappa1}),
(\ref{eq:kappa2}) and (\ref{eq:kappa4}) will be labeled by symbols
``$\kappa$'', ``$f^s$'', ``$\chi^{(4)}/\chi^2$'',
``$(\kappa^{(1)})^2/\chi$'', ``$\kappa^{(2)}/\chi$'' and
``$\kappa^{(4)}/\chi^2$'', respectively.

In table \ref{tab:2DI}, results for the critical coupling in
the Ising model on a square lattice are given.
The calculations were carried out for strips $L\times\infty$
with a periodic boundary condition in the transverse direction.
The estimates are shown for the pairs $(L-1,L)$ with $L\leq5$.
In the case of $(3,4)$ pairs, the errors are also given.
The type of phenomenological RG equations which have been used
are indicated in the first column of the table.
In this model $K_c={1\over2}\ln(1+\sqrt2)$
and $f_\infty=2G/\pi + {1\over2}\ln2$ ($G$ is Catalan's constant)
\cite{O44}.

\begin{table}
\caption{
Estimates of $K_c$ for the 2D sq Ising lattice;
$K_c^{exact}=0.440\,686\ldots$}
\label{tab:2DI}
\begin{tabular}{clll}
\hline
eq.&$(2,3)$&$(3,4)$&$(4,5)$\\[2mm]
\hline
 $(\kappa)$             &0.42236 &0.43088 $(-2.23\%)$ &0.43595\\
 $(\chi^{(4)}/\chi^2)$  &0.42593 &0.43242 $(-1.88\%)$ &0.43672\\
 $(\kappa^{(4)}/\chi^2)$&0.42596 &0.43243 $(-1.87\%)$ &0.43673\\[1mm]
 $(f^{s})$              &0.44324 &0.44168 $(+0.23\%)$ &0.44105\\
 $(\kappa^{(2)}/\chi)$  &0.47420 &0.45153 $(+2.64\%)$ &0.44626\\
\hline
\end{tabular}
\end{table}

It is seen from table \ref{tab:2DI} that the best lower bound
is given by eq.(\ref{eq:kappa4}).
Slightly worse results are obtained by Binder's
phenomenological renormalization-group procedure.
This approach which is normally implemented by Monte
Carlo simulations was used in the transfer-matrix version in
\cite{SD85}.
Nightingale's renormalization (first line in table 1) which is
traditionally used by transfer-matrix calculations has only the
third position in accuracy among the lower estimates.

I also found the phenomenological RG equations leading to the upper
bounds for $K_c$ (last two lines in table \ref{tab:2DI}).
Among these more accurate results are provided by eq.(\ref{eq:fs}).
The magnitude of the error in line four of table 1 is the least
among all lower and upper estimates of $K_c$.
Unfortunately, such approach requires a knowledge about the background
$f_\infty$.

Transfer-matrix eigenvalues for the Ising strips are known in
analytical form \cite{O44}.
Using this fact I considered the RG transformation
(\ref{eq:f^smn}) which makes use of the first derivative with
respect to $K$ $(m=1, n=0; h=0)$.
For the fixed point this transformation gives
\begin{equation}
   \label{eq:us}
   L^{d-y_t}u_L^s(K_c)=(L-1)^{d-y_t}u_{L-1}^s(K_c),
\end{equation}
where $u_L^s=u_L-u_\infty$ is the singular part of the reduced energy
density, and $u_L=\partial f_L/\partial K$.
Remarkably, the root of eq.(\ref{eq:us}) is equal to the exact
value of $K_c$ since $u_L^s(K_c)\equiv0$ for all $L$.
In other words, all finite-size corrections to the background
$u_\infty(=\sqrt2$ \cite{O44}) are zero.

Moreover, in the 2D Ising model $\partial\kappa_L/\partial K$ at
$K=K_c$ is also independed of $L$ and, therefore,
eq.(\ref{eq:kappa^mn}) with $m=1$ and $n=0$ gives
the exact value for the critical exponent: $\nu\equiv1/y_t=1$.

The data obtained for the 3-state square Potts lattice are
collected in table \ref{tab:2DP}.
For this model $K_c=\ln(1+\sqrt3)$ and
$f_\infty=4G/3\pi+\ln(2\sqrt3)+{1\over3}\ln(2+\sqrt3)$
\cite{B82}.
Inspecting table \ref{tab:2DP}, it is seen that eqs.(\ref{eq:fs})
and (\ref{eq:kappa1}) lead to more qualitative estimates than
Nightingale's approach.
Again, the lowest absolute error is yielded by the phenomenological
RG equation based on $f_L^s$.

Numerical calculations on strips $L\times\infty$ show us that in the
2D $q$-state Potts model the finite-size corrections to the background
energy, $u_\infty=1+1/\sqrt q$ \cite{B82}, are also absent and,
consequently, the equation
\begin{equation}
   \label{eq:u_L}
   u_L(K_c)=u_{L'}(K_c)
\end{equation}
yields the exact value of $K_c$.
Note that this equation has been derived earlier from other
considerations \cite{W94}.
\begin{table}
\caption{
Estimates of $K_c$ for the 2D sq 3-state Potts lattice;
$K_c^{exact}=1.005\,052\ldots$}
\label{tab:2DP}
\begin{tabular}{clll}
\hline
eq.&$(2,3)$&$(3,4)$&$(4,5)$\\[2mm]
\hline
 $(\kappa)$               &0.96248 &0.98350 $(-2.1\%)$ &0.99467\\
 $(\kappa^{(1)^2}/\chi)$&0.99311 &0.99920 $(-0.6\%)$ &1.00380\\[1mm]
 $(f^{s})$                &1.00927 &1.00667 $(+0.2\%)$ &1.00565\\
\hline
\end{tabular}
\end{table}

Let us discuss now the results presented in table \ref{tab:3DI} for
the 3D Ising model on a simple-cubic lattice.
For this model $K_c=0.221\,6544(3)$ \cite{TB96} and
$f_\infty=0.777\,90(2)$ \cite{M89}.
Renormalizations were done for the $L\times L\times\infty$
parallelepipeds with periodic boundary conditions in both
transverse directions.
As in the 2D case, the best lower values of $K_c$ are obtained
from eq.(\ref{eq:kappa4}).

\begin{table}
\caption{
Estimates of $K_c$ for the 3D sc Ising lattice;
$K_c^{exact}=0.221\,6544(6)$}
\label{tab:3DI}
\begin{tabular}{cll}
\hline
eq.&$(2,3)$&$(3,4)$\\[2mm]
\hline
 $(\kappa)$             &0.21340 &0.21826 $(-1.53\%)$\\
 $(\chi^{(4)}/\chi^2)$  &0.21823 &0.22002 $(-0.74\%)$\\
 $(\kappa^{(4)}/\chi^2)$&0.21824 &0.22006 $(-0.72\%)$\\[1mm]
 $(f^{s})$              &0.22354 &0.22236 $(+0.32\%)$\\
 $(\kappa^{(2)}/\chi)$  &0.22658 &0.22314 $(+0.67\%)$\\
\hline
\end{tabular}
\end{table}

In the 3D case the amplitudes of the finite-size
corrections to the critical-point energy are not equal to zero.
As a result, eq.(\ref{eq:u_L}) only yields an approximate
value of $K_c$.



The author thanks A.~A.~Belavin for stimulating discussions and
valuable remarks.
I am indebted also to M.~I.~Polikarpov for his encouragement.
This work was supported by RFBR Grant Nos.~99-02-16472 and
99-02-26660.



\end{document}